\RequirePackage{fix-cm}
\documentclass[smallextended,final,epjc3]{svjour3}          
\RequirePackage[T1]{fontenc}
\RequirePackage{graphicx}
\RequirePackage{lineno}
\RequirePackage{mathptmx}      
\usepackage{color}
\usepackage[usenames, dvipsnames, svgnames, table]{xcolor}

\RequirePackage[numbers,sort&compress]{natbib}
\RequirePackage[colorlinks,citecolor=blue,urlcolor=blue,linkcolor=blue]{hyperref}
\journalname{Eur. Phys. J. B}

\def\makeheadbox{{
\vspace{1.05cm}
\hbox to0pt{\vbox{\baselineskip=10dd\hrule\hbox
to\hsize{\vrule\kern3pt\vbox{\kern3pt
\hbox{\bfseries\small{ Eur. Phys. J. B}}
\hbox{\small{This is a post-peer-review, pre-copyedit version of this article.}}

\hbox{\small{The final version is available online at:} \href{https://doi.org/10.1140/epjb/e2019-100157-3}{https://doi.org/10.1140/epjb/e2019-100157-3}.}
\kern3pt}\hfil\kern3pt\hfil\kern3pt\vrule}\hrule}%
\hss}}}

\begin{document}

\title{Studies of Electronic Structure across a Quantum Phase Transition in CeRhSb$_{1-x}$Sn$_x$}


\author{R. Kurleto\thanksref{addr1}~\and~J.~Goraus\thanksref{addr2}~\and~M.~Rosmus\thanksref{addr1}~\and~A.~{\'S}lebarski\thanksref{addr2}~\and~P.~Starowicz\thanksref{e1,addr1}}

\thankstext{e1}{e-mail: pawel.starowicz@uj.edu.pl}

\institute{{Marian Smoluchowski Institute of Physics, Jagiellonian University, {\L}ojasiewicza~11, 30-348 Krak{\'o}w, Poland\label{addr1}} \and  \label{addr2}{Institute of Physics, University of Silesia, ul. 75 Pu{\l}ku Piechoty 1A, 41-500 Chorz{\'o}w, Poland}}

\date{Published in Eur. Phys. J. B, September 2019}
\maketitle
\begin{abstract}
We study an electronic structure of CeRhSb$_{1-x}$Sn$_x$ system, which displays quantum critical transition from a Kondo insulator to a non-Fermi liquid at~$x=0.13$. We provide ultraviolet photoelectron spectra of valence band obtained at 12.5~K. A~coherent peak at the Fermi level is not present in the data, but a signal related to~4f$^1$$_{7/2}$ final state is detected. Spectral intensity at the Fermi edge has a general tendency to grow with Sn~content. Theoretical calculations of band structure are realized with full-potential local-orbital minimum-basis code using scalar relativistic and full relativistic approach. The calculations reveal a depletion of density of states at the Fermi level for CeRhSb. This gap is shifted above the Fermi energy with increasing Sn content and thus a rise of density of states at the Fermi level is reflected in the calculations. It agrees with metallic properties of compounds with larger~$x$. The calculations also yield another important effects of Sn~substitution. Band~structure is displaced in a direction corresponding to hole doping, although with deviations from a rigid band shift scenario. Lifshitz transitions modify a topology of the Fermi surface a few times and a number of bands crossing the Fermi level increases. 
\end{abstract}

\section{Introduction}
	Quantum phase transition~(QPT) is a matter of particular interest because it is related to an instability of a ground state~\cite{hertz,sachdev,Coleman2005}. Heavy fermions are a perfect playground for studying quantum criticality. The transition from one to another type of the ground state is associated with non-Fermi liquid behavior. Quantum critical point~(QCP) which separates metallic phase from so called Kondo insulator~(KI) has attracted particular attention so far~\cite{si,diagram,prl2005}. KIs are materials with a narrow gap or a pseudogap in an electronic structure, which opens due to hybridization between correlated electrons and conduction band~\cite{psr,iga,prl2005}. 
		
	CeRhSb$_{1-x}$Sn$_x$ is an example of the system with QCP. Representatives of CeRhSb$_{1-x}$Sn$_x$ system crystallize in orthorombic~$\varepsilon$-TiNiSi structure (Pnma space group) for~$x<0.2$ and in hexagonal Fe$_2$P (P62m space group) for~$x>0.8$~\cite{prb2005,janka,pottgen}. In case of $0.2\le x\le0.8$ phase separation occurs. The parent compound, CeRhSb, is a KI with a gap width equal to approximately~7~K~\cite{iga,prb2005,malik91}. Doping with Sn, which is equivalent to increasing concentration of holes, leads to a gradual closing of the gap~\cite{sl2004}. The gap disappears at~$x=x_c=0.13$, which is associated with~QCP~\cite{prl2005}. For~$x>x_c$ the system becomes metallic with non-Fermi liquid ground state~\cite{prb2005,sl2006}. Thorough studies of physical properties of  CeRhSb$_{1-x}$Sn$_x$ in  the vicinity of the transition have been made so far~\cite{iga,prl2005,malik91}. According to them a universal scaling law~$\rho\cdot\chi=$const ($\rho$ - electrical resistivity, $\chi$ - magnetic susceptibility) is obeyed in KI phase~($x<0.13$)~\cite{prb2005}. A singular behavior of magnetic susceptibility was observed for~$x>x_c$, which is related to a non-Fermi liquid formation~\cite{sl2006}. At a higher value of temperature there is a crossover from a non-Fermi-liquid to a Fermi liquid~\cite{prl2005}. Strong valence fluctuations have been found in all synthesized representatives of CeRhSb$_{1-x}$Sn$_x$ system~\cite{malik91}. Such instabilities of Ce~valency are believed to play an important role in the pseudogap formation for~$x<x_c$~\cite{acta2000}.
	
	Photoelectron spectroscopy~(PS) combined together with theoretical modeling of a band structure is a powerful tool in studying effects of hybridization in heavy electrons systems. Several studies devoted to the electronic structure of~KIs have been conducted with application of PS, so far~\cite{breuer97,susaki96,breuer98,klein}. Particularly, the abrupt step behavior of the single ion Kondo scale~($T_K$) at a critical point has been inferred from photoemission data~\cite{klein}. 
	
	Both, CeRhSb and CeRhSn compounds were studied with application of PS. In case of CeRhSn~\cite{phmag1,sh}, an enhancement of signal attributed to 4f$^1$$_{5/2}$ final state, together with a small intensity of 4f$^0$ feature was interpreted as a symptom of valence fluctuations~\cite{phmag1,sh}. Valence band (VB) spectra of CeRhSb obtained previously show not only characteristics of a cerium compound with Kondo effect: f$^{0}$, f$^{1}_{5/2}$ and f$^{1}_{7/2}$ features but also a depletion of the spectral weight at $\varepsilon_F$ due to the hybridization gap formation below 120 K~\cite{tan,sh2,kumigashira99,kumigashira2001}. 
	
	Transition from KI in CeRhSb to a metallic state in CeRhSn via the critical point was studied by means of density functional theory~(DFT) calculations before~\cite{goraus2013}. It was shown that QPT is manifested by a change of Mulliken occupation of Ce~5d states as a function of $x$. Indeed, DFT calculations properly reproduced experimental value of critical Sn concentration~($x_c$=0.13) and it was testified that atomic disorder is not a necessary factor for the formation of~QCP. However, such a disorder leads to the Griffiths phase, which was evidenced for CeRhSn~\cite{sl2004,sl2006,gamza2009,phmag1}. 
		The schematic phase diagram of CeRhSb$_{1-x}$Sn$_x$ system in $T-x$ plane has been proposed so far~\cite{prb2005}. The metallization of a system was explained in terms of the model, which takes into account both, itinerant character of Ce~4f electrons and collective Kondo singlet state. A transition from KI to metallic state is related to destruction of such a collective singlet state accompanied by delocalization of 4f electrons.
		
	In this paper we present studies of the electronic structure of CeRhSb$_{1-x}$Sn$_x$ performed as a function of hole doping~$x$ in a light of~QPT. VB spectra collected with application of PS are confronted with the results of theoretical modelling. DFT~calculations show that the increase of~Sn content in CeRhSb$_{1-x}$Sn$_x$ results in a growth of density of states at the Fermi energy~($\varepsilon_F$), which is observed also in the experiment. The theoretical results also yield a sequence of Lifshitz transitions in Fermi surface and band structure.

\section{Material and methods}
	Description of synthesis and characterization of polycrystalline samples of CeRhSb$_{1-x}$Sn$_x$ ($x=0$, 0.06, 0.16, 1) was provided elsewhere~\cite{sl2004,prl2005}. The quality of used samples was checked with application of x-ray diffraction and energy dispersive x-ray spectroscopy~(EDXS). Diffraction patterns testified that the specimens subjected to the scrutiny consist of single phase. Good homogenity of samples was proven by means of~EDXS. The ultraviolet photoemission spectroscopy~(UPS) data have been collected with application of a VG~Scienta R4000 photoelectron energy analyzer. Measurements have been conducted at temperature equal to~12.5~K.  Both, He~II (40.8 eV) and He~I (21.2~eV) radiation have been used. Base pressure in the analysis chamber was equal to~4$\cdot$10$^{-11}$~mbar. Specimens were cleaned with diamond file in the preparation chamber (base pressure~4$\cdot$10$^{-10}$~mbar) prior to the measurement. For comparison we also made measurements on samples which were cleaved in the ultra-high vacuum conditions, but results obtained in this way have got worse quality. Overall resolution was not greater than~30~meV.
	
	Partial densities of states (DOSes) have been simulated with application of full-potential local-orbital (FPLO) package~\cite{koepernik99,opahle99,science2016,jssch2003,E32,ksiazka1}. Calculations in scalar relativistic scheme were performed with the virtual crystal approximation. Irreducible Brillouin zone (BZ) was divided for~343~k-points. Perdew-Wang exchange correlation potential was used. Partial DOSes, band structure along high symmetry directions in the first~BZ and Fermi surfaces have been resolved also with application of FPLO code in full relativistic approach.

	Photoelectron spectra have been simulated using calculated partial DOSes and assuming atomic cross sections for photoionization~\cite{yeh93}. Experimental broadening has been introduced by a convolution with the gaussian function of the width of~10~meV. 

\section{Results and Discussion}

\subsection{Photoemission - wide energy range}

	Spectra of CeRhSb$_{1-x}$Sn$_x$ measured in a wide range of binding energy ($\varepsilon_B$) are presented in~\hyperref[fig:vb]{Fig.~\ref*{fig:vb}~a}.  Sn~4d$_{5/2}$ and 4d$_{3/2}$ core levels have been observed at $\varepsilon_B$ equal to~$-24.74$~eV and~$-23.7$~eV, respectively, for compounds with $x\neq 0$, in He II spectra. Ce~5p doublet is visible for each investigated specimen. It consists of a humped structure at $\varepsilon_B\approx-21$~eV, followed by a broad peak at~$-17.9$~eV. This last structure was used for normalization of the He~II spectra, together with a guide served by Sn~4d core levels. A broad peak-like feature has been observed at $\varepsilon_B\approx-10$~eV except for the compound without~Sb. Signal in this energy region observed for CeRhSn is completely flat, so we can attribute this structure to Sb derived states. In fact, the position of the maximum corresponds to the Sb~5s states.

\subsection{Photoemission - valence band}	

\begin{figure} 
\begin{minipage}{\columnwidth}
\centering
\includegraphics[width=1\linewidth]{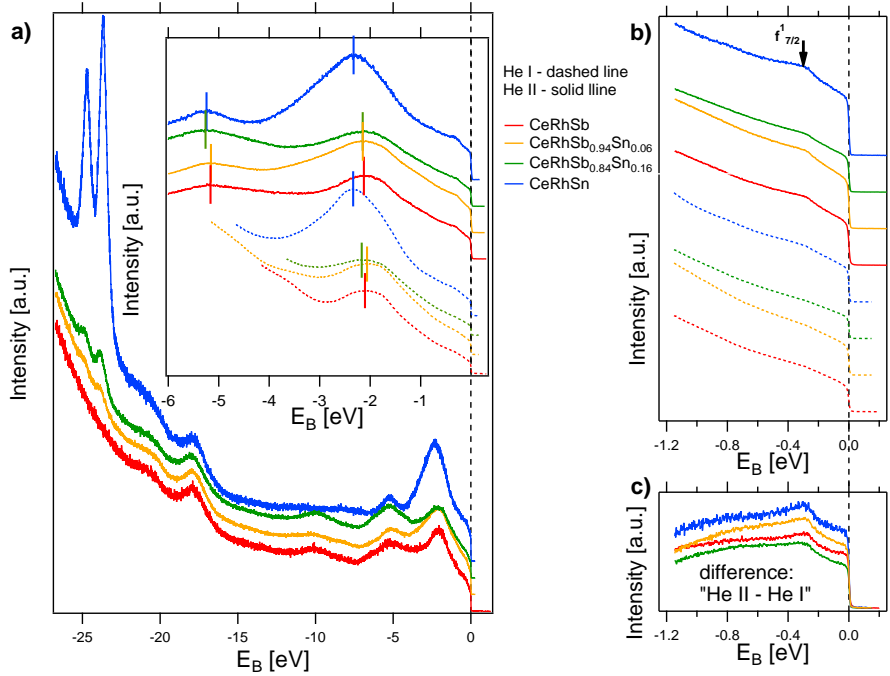}
\end{minipage}
\caption{(Color online) Valence band of the CeRhSb$_{1-x}$Sn$_x$ ($x=0$,~0.06,~0.16,~1) system studied by means of ultraviolet photoemission spectroscopy.  All measurements have been performed at T=12.5~K~a)~Spectra obtained with application of He~II radiation in wide energy range. (Inset)~Comparison of valence band spectra of studied compounds obtained with He~I and He~II radiation. Solid vertical lines denote peak maxima obtained with fitting the Gaussian function. b)~High resolution spectra of valence band near~$\varepsilon_F$, obtained with He~I and He~II radiation. c) Extracted 4f~contribution to spectral function for each studied compound.}
\label{fig:vb}
\end{figure}

	Spectra of investigated compounds have got similar shape in VB region~(inset~to~\hyperref[fig:vb]{Fig.~\ref*{fig:vb}~a}). There are two, broad and intense humps ($\varepsilon_B \approx -5.2$~eV,~$-2.1$~eV) and a small blurred peak close to $\varepsilon_F$ in He II measurements. The last structure ($\varepsilon_B\approx-0.3$~eV) is related to Ce~4f electrons, because it is suppressed in He~I spectra (discussed below). Locations and mutual ratio of intensities of ascribed structures depend on $x$. The peak corresponding to higher absolute value of~$\varepsilon_B$ is located at about~$-5.16$~eV in case of CeRhSb and remains in the same position for CeRhSb$_{0.94}$Sn$_{0.06}$, while in case of CeRhSn and CeRhSb$_{0.84}$Sn$_{0.16}$ it is found at about~$-5.25$~eV. This effect may be connected with a small contribution from Sn~p and d~states. The second structure also displays a weak dependency on~$x$. This maximum occupies roughly the same position, equal to~$-2.1$~eV, for three samples with~Sb, while for CeRhSn it moves to~$-2.3$~eV. Intensity taken at maximum at $\approx -2.1$ eV divided by intensity taken at second maximum amounts to~1.1,~1.3,~0.98 and~1.8 for~$x=0$,~0.06,~0.16 and~1 respectively. The values obtained for CeRhSn differ from these obtained for other studied compounds, which is not startling, because CeRhSn adopts a different crystal structure. Hence, it should have a different electronic structure as well. 
	
	Shape of VB spectra obtained with application of He~I radiation~(also shown in the inset~to~\hyperref[fig:vb]{Fig.~\ref*{fig:vb}~a}) corresponds to the data recorded with He~II lamp with two major peaks observed at similar binding energies. The spectrum is mainly related to~Rh~4d~states, what is supported by theoretical calculations discussed further in the text. This is also consistent with previous PS measurements and theoretical calculations performed for metallic~Rh~\cite{kelly84,smith74}. The maximum for CeRhSb is located at~$-2.1$~eV and a small amount of the dopant~($x=0.06$) shifts the peak towards~$\varepsilon_F$, next for~$x=0.16$ the maximum is displaced in opposite direction and finally for CeRhSn its binding energy is~$-2.33$~eV. These results would suggest a non-monotonic dependence with respect to~$x$. However, these displacements are very close to experimental uncertainty. He-II spectra do not follow well these trends.

	High resolution measurements of vicinity of~$\varepsilon_F$ are shown in~\hyperref[fig:vb]{Fig.~\ref*{fig:vb}~b}. Spectra measured with application of He~II are compared with those obtained with He~I. In He~II data one can see a broad maximum at about~$-0.3$~eV which is superimposed on a linear signal. The slope of measured intensity of He~I spectra changes as well. However, for He~I excited photoemission a broad hump near~$-0.3$~eV is visible only for CeRhSn. 
	
\subsection{Ce 4f spectral function}	
	
	Cross section for photoionization of Ce~4f electrons is amplified over three times when one goes from He~I to He~II~\cite{yeh85}, so we can estimate~4f contribution to spectral function by subtracting high resolution measurements: He~I from He~II~(\hyperref[fig:vb]{Fig.~\ref*{fig:vb}~c}). The obtained spectral function has got similar shape for each value of~$x$. A blurred peak-like maximum is observed roughly at the same~$\varepsilon_B=-0.3$~eV for each composition. This feature can be assigned to~4f$^1$$_{7/2}$ final state~\cite{sekiyama}.
	
	  The estimated Ce~4f spectral function of CeRhSb$_{1-x}$Sn$_x$ does not reveal any enhancement of the intensity at~$\varepsilon_F$ even for evidently metallic CeRhSn compound. This stands in a difference with some previous PES experiments performed on Kondo insulators, which often yield spectra which cannot be distinguished from  that of weakly hybridized Kondo metal~\cite{breuer98}, because of low experimental resolution or presence of nonmagnetic impurities.  Additionally, the lack of enhancement of signal at~$\varepsilon_F$ can be explained as follows. Firstly, theoretical calculations, which are discussed below, predict that Ce~4f partial DOS concentrates mostly above $\varepsilon_F$ for all possible concentrations of~Sn. Secondly, if the so called Kondo peak is present in spectral function of CeRhSn it is located, according to the Friedel sum rule, above $\varepsilon_F$~\cite{georges2016}. PES probes spectral function within occupied states, making observation of any enhancement of DOS almost infeasible in both situations.
	
	However, we believe that due to careful normalization of the spectra we can discuss how the spectral intensity at~$\varepsilon_F$ changes with substitution of Sn in place of Sb. Namely, one can see that for~$x=0.06$ the signal at~$\varepsilon_F$ increases by~13\%, while in case of~$x=0.16$, spectral intensity at $\varepsilon_F$ decreases by 16\%,  with respect to~$x=0$. Intensity measured at~$\varepsilon_F$ for CeRhSn is greater by about 33\% than that for CeRhSb. Nevertheless, samples with~$x=0.16$ are likely to be more defected, which may explain the lower signal at~$\varepsilon_F$. Hence, we cannot conclude, if a reduction of DOS for~$x=0.16$ is intrinsic to this~QPT or rather is an effect of disorder. The remaing compositions show a general increase of DOS at~$\varepsilon_F$ with doping.

\subsection{Theoretical calculations - scalar relativistic approach}
	
\begin{figure} 
\begin{minipage}{\columnwidth}
\centering
\includegraphics[width=0.6\linewidth]{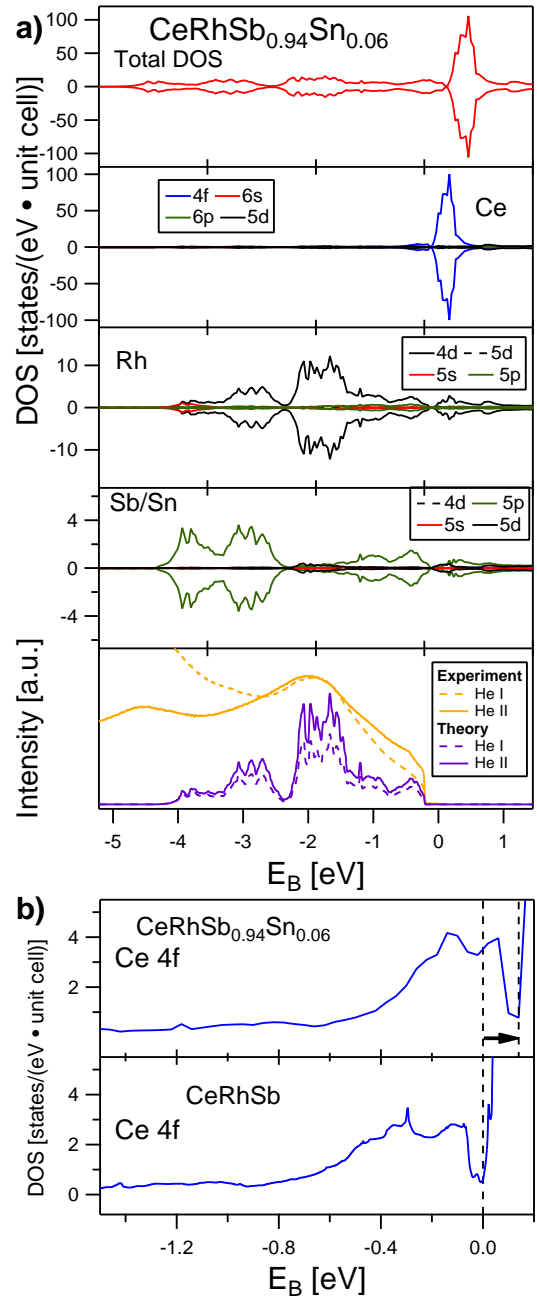}
\end{minipage}
\caption{(Color online) Results of~FPLO calculations with scalar relativistic approach. a)~Partial densities of states calculated for CeRhSb$_{0.94}$Sn$_{0.06}$. The lowest panel presents a comparison between experimental (collected at $T=12.5$~K) and simulated spectra of CeRhSb$_{0.94}$Sn$_{0.06}$. b)~Ce~4f partial DOS calculated for CeRhSb$_{0.94}$Sn$_{0.06}$ (upper panel) and for CeRhSb (lower panel). The black arrow indicates the position of the minimum in DOS which is discussed in the text. }
\label{fig:dos_sr}
\end{figure}

	 Total and partial DOS of CeRhSb$_{1-x}$Sn$_x$ have been calculated within scalar relativistic scheme for $0\le x\le0.3$. Calculated DOS has got a similar shape for each composition. We present results for CeRhSb$_{0.94}$Sn$_{0.06}$~(\hyperref[fig:dos_sr]{Fig.~\ref*{fig:dos_sr}~a}). VB is mainly composed of Ce~4f and Rh~4d states with admixture of Sn~5p and Ce~5d states. Partial DOSes are symmetric for both spin directions, as expected, because we have not included additional correlations in computational scheme. Ce~4f spectrum concentrates mainly above $\varepsilon_F$. Peak-like structure corresponding to 4f~states is centered at about~0.5~eV and it extends roughly from $\varepsilon_F$ to~0.9~eV. In the close vicinity of~$\varepsilon_F$ one can notice a depletion of Ce~4f~DOS~(\hyperref[fig:dos_sr]{Fig.~\ref*{fig:dos_sr}~b}). A "v-shape" gap is formed with non-zero DOS at minimum ($\approx$1~state/(eV$\cdot$unit~cell)). The calculations show that such a gap is located at~$\varepsilon_F$ for CeRhSb and is shifted above~$\varepsilon_F$ with Sn doping. As a consequence, DOS at~$\varepsilon_F$ is very low for a semimetallic CeRhSb and grows with~$x$. Such a gap is also present in Rh~4d and total~DOS. It is in line with ''semiconducting'' character of the system for~$0\le x\le x_c$. Such residual states in the gap are consistent with measurements of specif heat~\cite{nis96}. It is noteworthy, that Anderson lattice model at half filling also predicts a gap in 4f~spectral function, as well as in a spectrum of carriers from conduction band for KIs with momentum independent hybridization~\cite{grober98,rozenberg96}. 
	 	 
	 The  photoelectron spectra were simulated using theoretical DOS (cf. the lowest panel of~\hyperref[fig:dos_sr]{Fig.~\ref*{fig:dos_sr}~a}). They reproduce the shape of the valence band obtained experimentally. It appears that the contribution from Rh~4d  states dominates the valence band spectra. Rh~4d states extend mainly from~$\varepsilon_B$~equal to~$-5$~eV to about~2~eV. Rh~4d theoretical spectrum consists of two broad structures below $\varepsilon_F$, located at about~$-3$~eV and~$-2$~eV, respectively. Direct comparison of the simulated photoelectron spectra with the measurements shows good agreement in case of the structure at~$-2$~eV. However, the maximum located at~$-3$~eV in simulation is observed roughly at~$-5$ eV in the experimental data. 
	 
	 In~Sb/Sn spectrum we observe a peak originating from Sb~5s states, which is found approximately at~$-9$~eV~(not shown). This structure was also observed in He~II PS spectra near $\approx-10$~eV~(\hyperref[fig:vb]{Fig.~\ref*{fig:vb}~a}). A leading contribution from Sb/Sn site to DOS comes from 5p states. These states are smeared over whole VB, with increased contribution in $\varepsilon_B$ range from about~$-5$~eV~to~$-3$~eV. Sb/Sn~5p states are also depleted in the vicinity of~$\varepsilon_F$. 

\subsection{Theoretical calculations - full relativistic approach}
	
\begin{figure} 
\begin{minipage}{\columnwidth}
\centering
\includegraphics[width=1\linewidth]{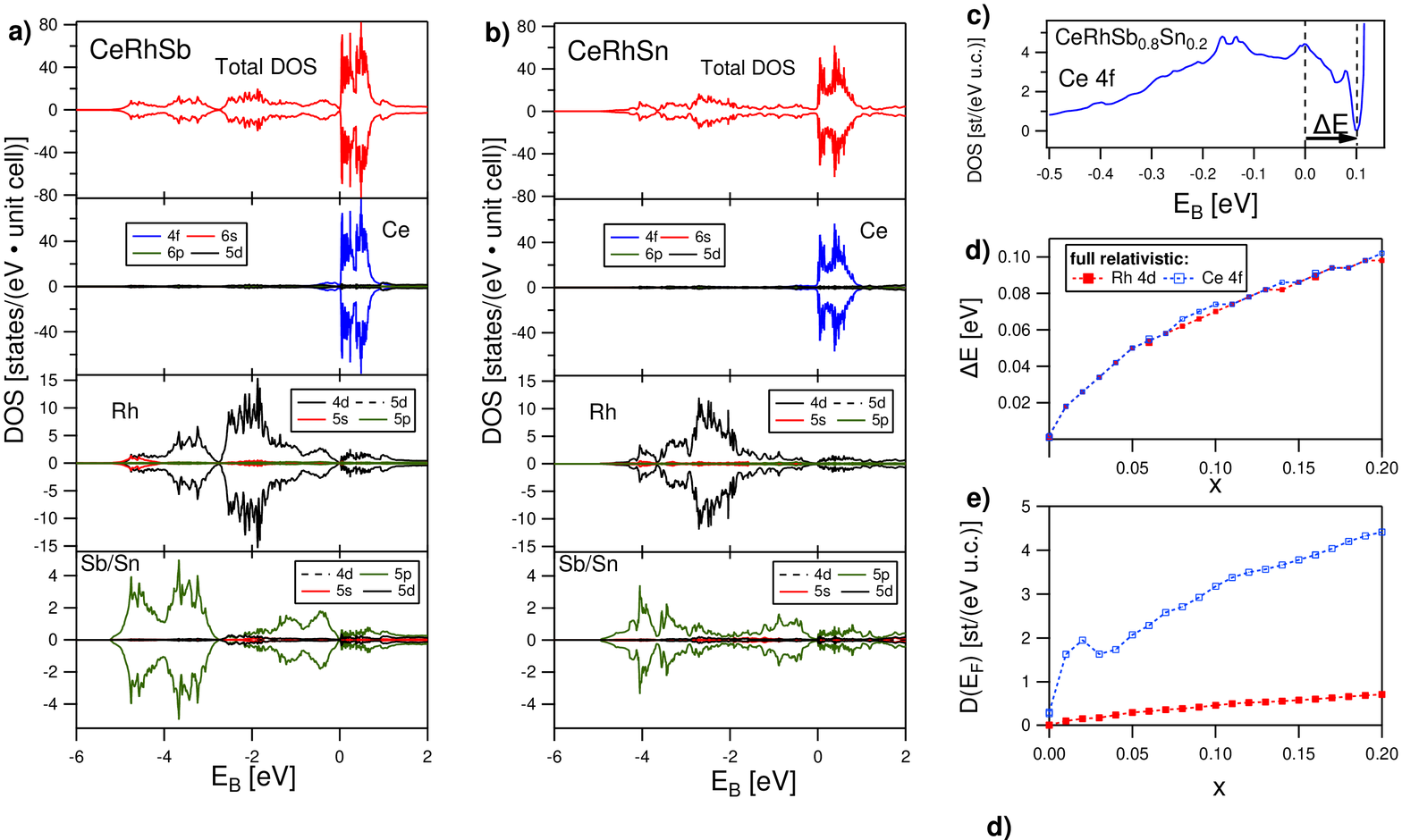}
\end{minipage}
\caption{(Color online) Results of FPLO calculations with full relativistic approach. Partial DOS is shown for (a) CeRhSb, (b) CeRhSn. (c)~Ce~4f partial DOS calculated for CeRhSb$_{0.8}$Sn$_{0.2}$. The black arrow indicates the position of the minimum in DOS, which is the closest to the Fermi energy. d)~Position of the partial Ce~4f and Rh~4d DOS minimum as a function of~$x$ in CeRhSb$_{1-x}$Sn$_x$. e)~Partial Ce~4f (blue open squares) and Rh~4d (red filled squares) densities of states at the Fermi level for CeRhSb$_{1-x}$Sn$_x$.}
\label{fig:graphs}
\end{figure}

	Partial DOS calculated in full relativistic approach for compounds with~$x=0$,~0.06, and~0.16, in principle has got a similar shape as this obtained within scalar relativistic scheme~(\hyperref[fig:graphs]{Fig.~\ref*{fig:graphs}~a,~b}). A difference is visible in Ce~4f spectra, the structure above~$\varepsilon_F$ is different than that  discussed previously. In case of~$x=0$,~0.06~and~0.16, the  gap-like depletion of~DOS in vicinity of~$\varepsilon_F$ is also observed~(\hyperref[fig:graphs]{Fig.~\ref*{fig:graphs}~c}).
	
	The position of the gap in Ce~4f and Rh~4d DOS is a rising function of~$x$~(\hyperref[fig:graphs]{Fig.~\ref*{fig:graphs}~d}). The gap is located at~$\varepsilon_F$ for CeRhSb. It is shifted above~$\varepsilon_F$ as a result of hole doping. The same trend is observed for Ce~4f and Rh~4d states. The influence of the doping is also visible in DOSes at~$\varepsilon_F$~(\hyperref[fig:graphs]{Fig.~\ref*{fig:graphs}~e}). The Rh~4d partial DOS at~~$\varepsilon_F$ rises almost linearly with~$x$, while in case of Ce~4f a small maximum at~$x=2$ is visible. However, both Ce~4f  and Rh~4d DOS at~$\varepsilon_F$ are enhanced with respect to~$x=0$.
	
	Partial DOS calculated for different orbital characters in CeRhSn, is different from the corresponding DOS in the compounds with non-zero Sb amount. There is no longer gap in vicinity of~$\varepsilon_F$. Strictly speaking, in Rh~4d and in Sn~5p spectral densities, there is a significant, gap-like depletion at $\varepsilon_F$. However in case of Ce~4f, $\varepsilon_F$ is located in the rising slope of the peak in DOS. One can conclude that metallic properties of CeRhSn are mostly related to 4f~electrons. 
	
\begin{figure} 
\begin{minipage}{\columnwidth}
\centering
\includegraphics[width=1\linewidth]{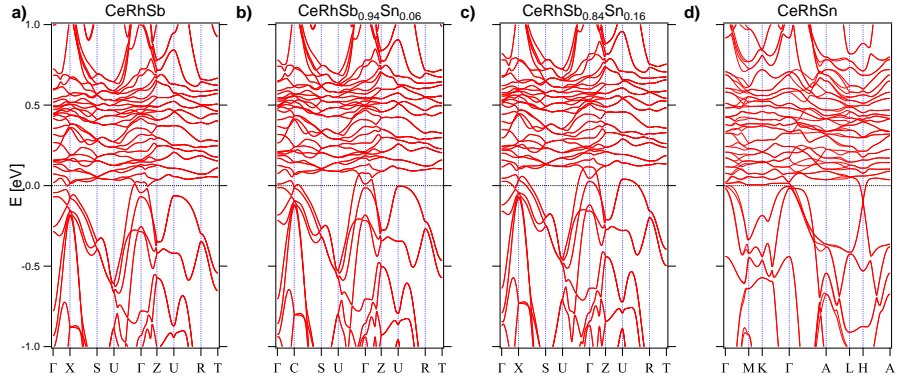}
\end{minipage}
\caption{(Color online)~Band structure along high symmetry directions calculated for CeRhSb$_{1-x}$Sn$_x$ ($x=0$,~0.06,~0.16,~1) with~FPLO method in full relativistic approach.}
\label{fig:bands}
\end{figure}

	Dispersion of the bands in the first BZ, calculated within full relativistic FPLO method, is shown in~\hyperref[fig:bands]{Fig.~\ref*{fig:bands}}. For all studied compounds, except for CeRhSn, bands have similar shape. At the first sight it seems that Fermi level is simply shifted in a direction corresponding to lowering electron band filling. However, a more detailed analysis shows that there are significant deviations from a rigid band shift. Nevertheless, due to displacing bands below Fermi energy and shifting another bands to the Fermi level one can anticipate interesting modifications of the Fermi surface topology.  The observed band shifts suggests that certain filled bands may become hole bands and some electron parts of the Fermi surface may disappear with doping. This would mean that Lifshitz transitions occur with growing~$x$. Careful analysis of the evolution of~FS may confirm that and it is presented in the next section.

	The bands related mostly to 4f~electrons are characterized by a very weak dispersion and lie just above~$\varepsilon_F$. In band structure of CeRhSn there are bands located nearly at~$\varepsilon_F$ related to 4f~electrons and characterized by high effective mass. There is one band along M-K-$\Gamma$ direction, which is especially flat and lies just above~$\varepsilon_F$.

\subsection{Calculated Fermi surfaces}	
	
\begin{figure} 
\begin{minipage}{\columnwidth}
\centering
\includegraphics[width=0.6\linewidth]{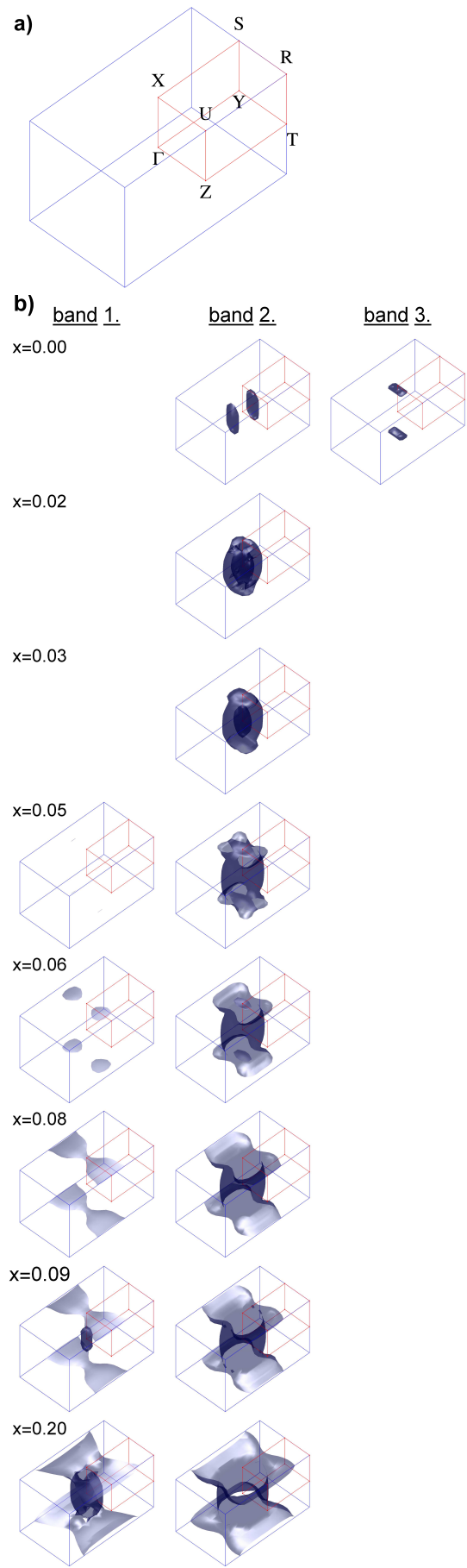}
\end{minipage}
\caption{(Color online)~a)~The first Brillouin zone of CeRhSb$_{1-x}$Sn$_x$. b)~The Fermi surface of selected representatives of CeRhSb$_{1-x}$Sn$_x$ system.}
\label{fig:FS}
\end{figure}

		Theoretical calculations performed in full relativistic approach have yielded Fermi surfaces (FSes) of representatives of CeRhSb$_{1-x}$Sn$_x$ family~(\hyperref[fig:FS]{Fig.~\ref*{fig:FS}}). The shape of the FS strongly depends on~$x$. The  band structure calculated along high symmetry lines~(\hyperref[fig:bands]{Fig.~\ref*{fig:bands}}) shows that the number of bands crossing $\varepsilon_F$ depends on the sample composition. The same feature is reflected in the FS contours. The shape of~FS changes with~$x$ in a discontinuous manner. The variation of the number of separated parts of~FS foreshadows the topological changes of FS, which can be regarded as a sequence of subsequent Lifshitz transitions. In case of CeRhSb, four hole-like pockets contribute to the~FS. They are transformed under doping~($x=0.02$) into one tori-like structure extended along~$\Gamma$-X~direction. For~$x=0.03$ FS can be regarded as two deformed concentric spheres. The increase in $x$ leads to increase of the overall volume of FS. For~$x=0.05$ FS contour reaches the border of the BZ, resulting in opening of FS. Moreover, some new parts of~FS, attributed to band~1 emerge. In case of~$x=0.6$, FS consists of four leaf-like structures near the~X-U-S~BZ~wall (band~1) and a barrel-like shape along~$\Gamma$-X direction with two plate-like structures on both sides of the~X~point (band~2). Subsequent increase in~$x$ makes leaf like structures merge into a pair of bow-tie shaped structures. The barrel-like shape becomes bigger, while the bottoms (plate-like structures) disappear. Afterwards, the additional closed part of~FS appears around the~$\Gamma$ point, for~$x=0.09$. The volume of this structure increases with~$x$, and eventually it touches and merges with the bow tie-like structure near to~X-U-S wall of~BZ. In parallel, the volume of the barrel like band~2 increases.

		The presented simulations indicate that~Sn substitution in CeRhSb$_{1-x}$Sn$_x$ system modifies~FS topology six times for~$x$ ranging from~0~to~0.2. This is interpreted as a sequence of Lifshitz transitions. FS volume is also considerably increased. It is difficult to assign a particular Lifshitz transition to~QPT because the location of~$\varepsilon_F$ is given with some uncertainty in~DFT calculations. Hence, Sn concentrations in calculations may correspond to a shifted value of~$x$ in a real system. Nevertheless, a general increase of~DOS at $\varepsilon_F$, FS volume and a number of band crossing~$\varepsilon_F$ certainly helps to understand the modifications of electronic structure, which are related to QPT. 

\section{Conclusions}

	We have investigated electronic structure of CeRhSb$_{1-x}$Sn$_x$ family, in which~QPT occurs. VB~spectra obtained with~UPS agree qualitatively with both scalar relativistic and full relativistic~FPLO calculations. They are dominated by~Rh~4d~states. The spectra collected at~12.5~K, do not reveal 4~f electron peak in vicinity of~$\varepsilon_F$.  However, the feature related to~4f$^1$$_{7/2}$ final state is clearly visible in Ce~4f spectral function extracted for each of the studied compounds. The UPS results exhibit a tendency of Ce~4f derived~DOS~at~$\varepsilon_F$ to rise with Sn content. Calculated DOS yields a semimetallic character of CeRhSb with a gap-like minimum in~DOS~at~$\varepsilon_F$. This gap is shifted above~$\varepsilon_F$ with Sn doping leading to metallic properties. Full~relativistic calculations show that due to the effect of hole doping realized with Sn substitution a few modifications of~FS topology takes place. These are interpreted as Lifshitz~transitions. The calculations also yield that~FS volume generally grows and a number of band crossings at $\varepsilon_F$ increases as a function of~$x$. These results visualize the evolution of the electronic structure which occurs in~QPT in CeRhSb$_{1-x}$Sn$_x$ system.

\newpage

\begin{acknowledgements}
This work has been supported by the National Science Centre, Poland within the Grant no. 2016/23/N/ST3/02012. Support of the Polish Ministry of Science and Higher Education under the grant 7150/E-338/M/2018 is acknowledged.
\end{acknowledgements}

\section*{Author contribution statement}
A.~{\'S}. and P.~S. came up with the presented idea. A.~{\'S}. synthesized and characterized the used samples. J.~G. performed theoretical calculations. Photoelectron spectroscopy was performed by R.~K., M.~R. and P.~S. R.~K. performed data analysis and comparison with theoretical results.  The project was realized under supervision of P.~S. All authors discussed the obtained results and contributed to the article.

\bibliographystyle{spphys} 
\bibliography{mybibfile}

\end{document}